# Scheme for proton-driven plasma-wakefield acceleration of positively charged particles in a hollow plasma channel

Longqing Yi (易龙卿),[1] Baifei Shen (沈百飞),[1]* Konstantin Lotov,[2,3] Liangliang Ji (吉亮亮),[1] XiaomeiZhang (张晓梅),[1] Wenpeng Wang (王文鹏),[1] Xueyan Zhao (赵学燕),[1] Yahong Yu (郁亚红),[1] Jiancai Xu (徐建彩),[1] Xiaofeng Wang (王晓峰),[1] Yin Shi (时 银),[1] Lingang Zhang (张林港),[1] Tongjun Xu (徐同军),[1] and Zhizhan Xu (徐至展)[1]

[1] *State Key Laboratory of High Field Laser Physics, Shanghai Institute of Optics and Fine Mechanics, Chinese Academy of Sciences, P.O. Box 800-211, Shanghai 201800, China*

[2] *Budker Institute of Nuclear Physics, 630090, Novosibirsk, Russia*
[3] *Novosibirsk State University, 630090, Novosibirsk, Russia*

**Abstract**

A new scheme for accelerating positively charged particles in a plasma wakefield accelerator is proposed. If the proton drive beam propagates in a hollow plasma channel, and the beam radius is of order of the channel width, the space charge force of the driver causes charge separation at the channel wall, which helps to focus the positively charged witness bunch propagating along the beam axis. In the channel, the acceleration buckets for positively charged particles are much larger than in the blowout regime of the uniform plasma, and stable acceleration over long distances is possible. In addition, phasing of the witness with respect to the wave can be tuned by changing the radius of the channel to ensure the acceleration is optimal. Two dimensional simulations suggest that, for proton drivers likely available in future, positively charged particles can be stably accelerated over 1 km with the average acceleration gradient of 1.3 GeV/m.

The development of conventional accelerators has brought us to the brink of a new era of particle physics. High quality proton beams as energetic as 7 TeV are now available in Large Hadron Collider (LHC) at European Center for Particle Physics (CERN). The energy frontier for lepton accelerators is, however, much lower, since it takes many kilometers to generate particle energies of interest to high energy physicists. The reason is that the acceleration gradient in radiofrequency-based accelerators is limited to 20-50 MV/m as a result of material breakdown. Enormous expense and occupancy area put their constraints on future accelerators based on conventional methods.

On the other hand, plasma-based accelerators[1], which have ability of sustaining extremely large acceleration gradients, orders of magnitude higher than the breakdown fields of accelerators, have been making a remarkable progress in electron acceleration[2-8]. However, there is no such impressive progress for the positively charged particles[9]. One of reasons is that strongly nonlinear plasma wave is asymmetric with respect to the charge sign of accelerated particles. In the blowout regime, the region suitable for focusing and acceleration of positrons is extremely narrow[10, 11]. Moreover, the high density of plasma electrons in this region makes it impossible to achieve a low energy spread and conservation of the normalized emittance. An alternative advanced acceleration technique, the light pressure acceleration (LPA)[12-17] meets huge difficulties due to high requirements for the power and contrast of the laser pulse. Also, the unfavorable scaling of the energy gain with the acceleration distance, $\varepsilon \propto L^{1/3}$, prevents the proton energy from going beyond 100 GeV.

In order to break both of the two bottlenecks, we propose to use a plasma vacuum channel to overcome the problem of defocusing of the witness beam and sustain a large acceleration gradient at the same time. Unlike previous studies on drive particle beams propagating in a channel[18, 19], where dielectric wakefield accelerator (DWA)[20, 21] or phase mixing[19] is mostly concerned, the channel radius $r_0$ in our scheme is relatively small, usually of the same order as the radial rms length $\sigma_r$ of the driver, i.e. $r_0 \sim \sigma_r$. Therefore, the space charge force of the driver causes charge separation at the wall of channel, which helps to focus the witness bunch propagating along the beam axis as shown by the sketch map in Fig. 1. The witness beam located at the proper wakefield phase could gain a considerable energy. We show that specific problems of positively charged particle acceleration in nonlinear plasma waves can be solved with this scheme.

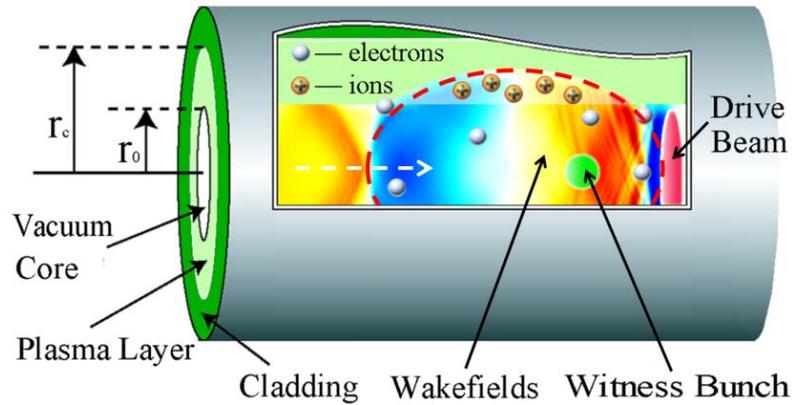

Fig. 1 (color online) Conceptual drawing of the scheme. The proton drive beam excites a plasma wave in the hollow plasma channel, while the positively-charged witness bunch gains energy from the longitudinal component of the wake (shown by the colors) and is focused by the electrons entered into the channel.

Plasma wakefield acceleration in channels is not free from transformer ratio limit[22] stating that the energy gain of the witness beam cannot in practice exceed roughly 2 times the driver energy. Since protons can carry much more energy than electrons in nowadays accelerators, they have the potential to excite the wakefiled over a long distance and accelerate a witness beam to very high energies[23]. In our scheme, a TeV proton beam is used as the driver to generate strong plasma wakefields, which in principle enables a witness bunch to gain TeV-range energies in a single stage of acceleration.

While positrons can certainly be accelerated with this scheme, here we report our simulation results of single-stage proton acceleration as the primary example. At high energies, protons behave similarly to positrons, but are easier to be simulated with the available code. We use a 2 TeV driving proton bunch and perform two dimensional (2d) quasi-static simulations with LCODE[24]. Other parameters of the driver are those used in Ref.[23]; the higher energy is taken just to obtain larger single-stage energy gain. The witness protons are injected 0.75 mm behind the driver; the total witness charge is 1 nC. The initial spot size and bunch length of the witness beam are 200 μm and 25 μm, respectively, and initial divergence is 0.03 mrad. This relatively small bunch length is used to reduce its unfavorable energy-spread broadening induced by the difference of the acceleration along beam axis. In our scheme, the long channel is consisted of many short capillaries which have been already used for laser electron acceleration. It should be mentioned that no discharge happens and there is no any plasma before the drive proton beam comes. Therefore, such hollow channels are very easy to be controlled. It is the drive proton beam pulls the electrons out of the channel wall and the plasma is formed in this way. The discharged channel is modeled by a plasma with constant density $n_0 = 6 \times 10^{14} \text{cm}^{-3}$ outside the plasma channel ($r \geq r_0$) and $n = 0$ inside the channel ($r < r_0$) in the simulation. We have scanned over a range of channel radii and found out that $r_0 = 0.64$ mm provides the best performance. We also use an axisymmetric analog of external focusing quadrupoles[25] to guide the very head of the proton driver, which otherwise undergoes a nature dispersion. Note that the focusing strength of the quadrupoles is too weak to have an influence on the witness beam. Detailed simulation parameters are listed in Table 1.

| Parameters | Symbols | Values | Units |

| Number of protons in drive beam | $N_p$ | $10^{11}$ | |
|---|---|---|---|
| Proton Energy | $E_p$ | 2 | TeV |
| Initial longitudinal size of drive beam | $\sigma_z$ | 100 | μm |
| Initial transverse size of drive beam | $\sigma_r$ | 0.43 | mm |
| Initial longitudinal momentum spread of drive beam | $\Delta P_z/P_z$ | 0.1 | |
| Initial angular spread of drive beam | $\Delta P_r/P_z$ | $3 \times 10^{-5}$ | |
| Total charge of the witness bunch | $Q_w$ | 1 | nC |
| Plasma density | $n_0$ | $6 \times 10^{14}$ | $cm^{-3}$ |
| Plasma channel radius | $r_0$ | 0.64 | mm |
| Magnetic field gradient | $S$ | 0.25 | T/mm |
| space period of the quadrupoles | $L_q$ | 3 | m |

Table 1 Parameters for the simulation.

Single stage acceleration over 1000 meters has been simulated. Figure 2(a-d) shows snapshots of the particle phase space at several distances along the channel. It is seen that during propagation of the drive proton bunch though the channel, its tail loses significant amounts of energy, while the witness beam picks up the energy. The acceleration process lasts for as long as 1000 meters until it stops due to the phase slippage effect. About 67.4% of the witness particles are remaining in the bunch over

1 km, i.e. the final charge is 0.674 nC. The witness bunch reaches the mean energy of 3.3 TeV after one acceleration stage. The final energy spread of the witness beam is about 2.6%.

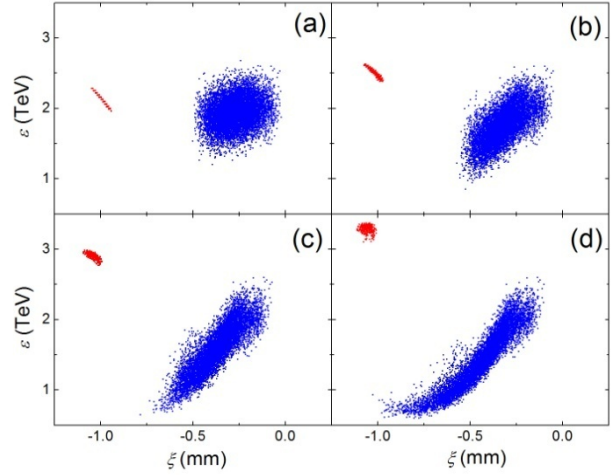

Fig. 2 (color online) Phase space (energy versus relative position) of the drive beam (blue dots) and witness bunch (red dots) taken at propagation distances of (a) 100 m, (b) 400 m, (c) 700 m, and (d) 1000 m.

The initial mean energy of witness protons is 2 TeV, and their energy is assumed to depend linearly on the longitudinal coordinate $z$, i.e. $\varepsilon_i(z) = \varepsilon_1(1 - z/\sigma_w) + \varepsilon_2 z/\sigma_w$, where $z$ is measured from the bunch head, $\sigma_w$ is the longitudinal size of the witness beam, $\varepsilon_1 = 1.8$ TeV and $\varepsilon_2 = 2.2$ TeV are the initial energies of the bunch head and tail, respectively. This energy distribution is chosen to compensate different acceleration gradients felt by different parts of the witness bunch as shown in Fig. 3. Unlike electrons, positively charged particles are accelerated in the front part of the "bubble", and forefront witness particles feel a stronger accelerating gradient. This positive feedback results in increase of the absolute energy spread if the initial energy is constant distributed along the bunch.

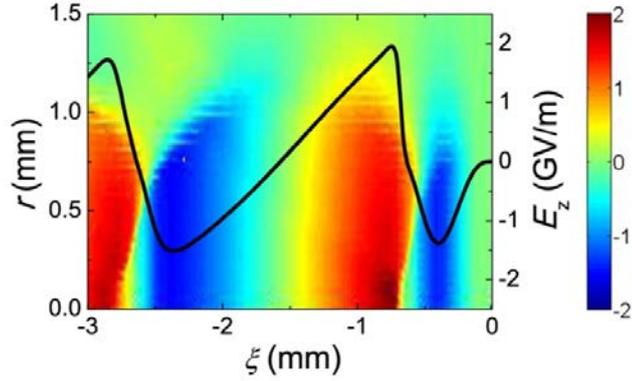

Fig. 3 (color online) Two dimensional distribution of the accelerating field generated by the proton bunch travelling in the plasma channel with radius $r_0 = 0.64$ mm. The on-axis field is shown by the black curve.

Compared to the uniform plasma case, the plasma channel does not change the accelerating field much. As is shown by Fig. 3, the amplitude of the wakefield generated by the proton beam is about 2 GV/m, slightly smaller than that in the uniform plasma of the same density (about 3 GV/m[23]). This is because the positively charged driver induces the phase mixing effect[18] in the uniform plasma, thus reducing the accelerating field as well. In contrast, in the plasma channel, the phase mixing is not so important[19], since there are no plasma electrons near the axis which oscillate in different phases as compared to electrons originating from far away from the axis. Besides, in our scheme both the amplitude and period of the plasma wakefield change as the radius of plasma channel changes. As $r_0$ becomes larger, the plasma wave amplitude decreases and the period increases. This effect offers a way to tune the acceleration phase of the witness bunch by changing the radius of the plasma channel.

At the cost of small decrease of the acceleration gradient, the radial field in our scenario (shown in Fig. 4(a) and (b)) is much improved compared to that in the uniform plasma case. The resulted focusing force is weak on the axis and nonlinear along radial distance, which remaining relatively constant along the acceleration distance, enabling a stable acceleration of witness bunch over thousands of meters.

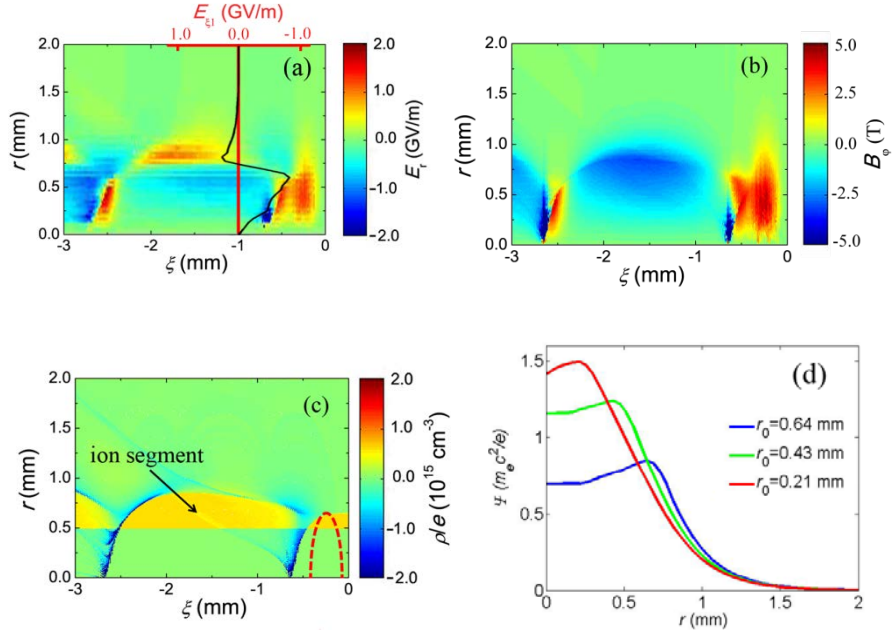

Fig. 4 (color online) (a) Simulated map of the radial electric field, (b) magnetic field, (c) net charge density, and (d) radial dependencies of the potential $\Psi(r) = -\int_r^\infty (E_r - B_\varphi) dr$ at the witness location (red vertical line in (a)) for several channel radii. Radial electric field $E_{\xi 1}(r)$ at the cross-section of witness bunch is shown by the black curve in (a). The red dashed curve in (c) shows the position of proton driver.

Figure 4(a) indicates that when the proton beam travels in the channel, the radial electric field $E_r$ in the excited "bubble" switches its sign at the channel boundary, so that it acts as a focusing force on positively charged particles. However, the magnetic field $B_\varphi$ shown in Fig. 4(b) is almost the same as in the uniform plasma case and tends to expel these particles out. The competition between $E_r$ and $B_\varphi$ determines the focusing properties of the channel.

In order to get the full picture, we plot in Fig. 4(d) the simulated radial distributions of the wakefield potential $\Psi(r) = -\int_r^\infty (E_r - B_\varphi)dr$ at $\xi = \xi_1$ for different channel radii. In all cases, there is always a potential well on the axis ($r = 0$), and the potential gradients near the beam axis are relatively small, which indicates a small but nonzero focusing force exists in this region. As a result, positively charged particles in the channel, i.e. at radii $r < r_0$, can be confined in a large acceleration bucket. Outside the channel ($r \geq r_0$), the force is defocusing and particles with too large transverse momentum will be dispersed. Besides, Fig. 4(d) also shows that both the depth and width of the potential well vary as the channel radius changes. As the channel radius grows, the potential well becomes deeper and wider so that confinement of the witness beam becomes better at the cost of slight decrease of the acceleration gradient.

Analysis on the focusing properties of the channel relies on the charge density distribution shown in Fig. 4(c). One can see the segment of positive charge is formed by the background ions. Plasma electrons from this segment are expelled out by the space charge force of the driver. Most of them enter the channel and concentrate in the front and back edges of the segment, while ions are left at initial positions because of their much heavier mass. Compared to the well-known "bubble" regime in the uniform plasma case, a hollow "bubble" is formed here, with no plasma ions inside the channel ($r < r_0$).

The radial electric field and the magnetic field in the "bubble" can be derived

from the Poisson and Maxwell equations

$$\frac{1}{r}\frac{\partial}{\partial r}(rE_r) = 4\pi e(n_b + n_i) - \frac{\partial E_z}{\partial \xi}, \tag{1}$$

$$\frac{1}{r}\frac{\partial}{\partial r}(rB_\varphi) = 4\pi e n_b - \frac{\partial E_z}{\partial \xi}, \tag{2}$$

Where $n_i$ is the density of plasma ions, $\xi = z - v_p t$, $n_b$ is the beam density of proton driver, $e$ is the fundamental charge, $v_p$ is the phase speed of the plasma wave, and we neglected the low density of plasma electrons. The Lorentz force acting on a witness proton is

$$F_r = e(E_r - B_\varphi) = \begin{cases} 2\pi n_i e^2 (r^2 - r_0^2)/r, & r \geq r_0 \\ 0, & r < r_0 \end{cases}. \tag{3}$$

One can see that $F_r$ has a defocusing effect for the positively charged particles outside the channel, which is responsible for the wakefield potential barrier near $r = r_0$ as shown in Fig. 4(d).

Equation (3) also indicates the electric and magnetic forces are completely cancelled inside the channel, which means there is neither defocusing force nor focusing force there. Fortunately, some electrons enter the channel as a result of the "sucked-in" effect[23] of the positively charged driver, forming a shallow potential well on the axis to provide a weak focusing force inside channel. as is shown in Fig. 4(d).

In the following, we will give a scaling relationship between the witness energy gain and driver parameters. After long-distance acceleration in the plasma channel, protons in the tail of the drive bunch lose significant energy, which results in a phase slippage between those decelerated protons and the plasma wave. As more and more driver protons fall into the acceleration phase of the plasma wave, the wave is eventually "overloaded" by these protons and the acceleration gradient decreases significantly. The relative slippage between the plasma wave and decelerated protons in the tail of the drive beam is[22]

$$\Delta L = \frac{L}{2}\left(\frac{1}{\gamma_{ti}\gamma_{tf}} - \frac{1}{\gamma_{pi}\gamma_{pf}}\right), \tag{4}$$

where $L$ is the dephasing length, $\gamma_p$ and $\gamma_t$ are Lorentz factors of the plasma wave and protons in the tail of the driving bunch. Subscripts "i" and "f" refer to initial and final values, respectively. For the proton driven plasma wakefield, $\gamma_{pi} = \gamma_{ti} = \gamma_0$ being the initial Lorentz factor of the driving beam. The phase velocity of the plasma wave changes little during the acceleration stage, i.e. $\gamma_{pf} \approx \gamma_0$. The dephasing distance can be written as

$$L = -\frac{2\gamma_0^2 \gamma_f}{\gamma_0 - \gamma_f}\Delta L. \tag{5}$$

Here we have used $\gamma_f$ instead of $\gamma_{tf}$ in Eq. (5) for simplification. Let $\bar{E}$ be the average decelerating electric field felt by the drive bunch tail, the energy equation is,

$$\gamma_f m_p c^2 - \gamma_0 m_p c^2 = \bar{E}eL, \tag{6}$$

where $m_p$ is the proton mass. Equations (5) and (6) give the total acceleration distance,

$$L = -\gamma_0^2 \left(\sqrt{1 + \frac{2m_p c^2}{\bar{E}e\Delta L}\frac{1}{\gamma_0}} - 1\right)\Delta L. \tag{7}$$

According to the simulation, relative slippage distance $\Delta L \approx -0.5$ mm and average decelerating field $\bar{E} \approx -1.5$ GV/m. For a proton driver with the initial energy 2 TeV ($\gamma_0 = 2134$), the calculated acceleration distance is $L \approx 1060$ m, while in our simulation the acceleration stops at about 1 km because of the phase slippage. Multiplying the acceleration distance by the average accelerating force acting on a witness particle, we can obtain the final energy gain of the witness $\varepsilon$ as a function of the initial Lorenz factor of the driving beam $\gamma_0$.

The value of $\Delta L$ and $\bar{E}$ are independent with $\gamma_0$, so one can see from Eq. (7) that the energy gain $\varepsilon$ scales as $\gamma_0^\alpha$, where $1 < \alpha < 1.5$ depends on the value of $\gamma_0$. The exponent $\alpha$ is close to 1.5 if $\gamma_0$ is relatively small and decreases as $\gamma_0$ grows. If $\gamma_0$ is large enough, i.e. $\gamma_0 \gg \frac{2m_\mathrm{p} c^2}{\bar{E} e \Delta L} \sim 2400$, $\varepsilon$ scales as $\gamma_0$ almost linearly ($\alpha \to 1$), thus giving a constant radio of the witness energy gain to the initial driver energy.

In conclusion, the simulation results indicate that a positively charged bunch could be stably accelerated in the nonlinear plasma wakefield excited by a high energy proton beam inside a narrow plasma channel of the radius $r_0 \sim \sigma_\mathrm{r}$. The acceleration gradient is as large as several GV/m, the accelerating bucket for positively charged witness is much bigger than that in the uniform plasma, and the radial focusing force in the acceleration bucket is small but nonzero. The confining effect originates from the charge separation at the channel wall. The acceleration gradient and the period of the plasma wave both change as the channel radius changes, which could be a way to tune the acceleration phase of the witness bunch to ensure the acceleration is optimal. Acceleration is limited by the phase slippage between decelerated driving protons and the plasma wave. As an example, proton acceleration over 1 km is simulated. With the proper (linear) initial energy distribution of the witness bunch, a high quality proton beam is obtained with the energy increased by 1.3 TeV and the final energy spread of about 2.6%. According to the obtained scaling, the energy gain in a single acceleration stage can be as high as 4.5 TeV if one uses a 7 TeV LHC beam as the driver.

This work has been supported by the Ministry of Science and Technology (2011CB808104, 2011DFA11300), National Natural Science Foundation of China (Projects No. 61008010, No. 11125526, No. 11127901, No. 10834008, and No. 60921004), and the Ministry of Education and Science of Russia (No. 14.B37.21.0784).

*To whom all correspondence should be addressed. bfshen@mail.shcnc.ac.cn.